\documentclass[prc,twocolumn,nofootinbib]{revtex4-2}
\usepackage{color}

\usepackage{amsmath}
\usepackage{amssymb}
\usepackage{graphicx}
\usepackage{bm}
\usepackage{hyperref}
\usepackage{url}
\hypersetup{colorlinks = true, allcolors = blue}

\begin{document}

\title{Applications of the dynamical generator coordinate method to quadrupole excitations}

\author{N. Hizawa}
\author{K. Hagino}
\author{K. Yoshida}
\affiliation{
Department of Physics, Kyoto University, Kyoto 606-8502,  Japan}
\begin{abstract}
We apply the dynamical generator coordinate method (DGCM) with a conjugate momentum to a nuclear collective excitation.
To this end, we first discuss how to construct a numerically workable scheme of the DGCM for a general one-body operator.
We then apply the DGCM to the quadrupole vibration of $^{16}$O using the Gogny D1S interaction. 
We show that both the ground state energy and the excitation energies are lowered as compared to the conventional GCM with the same number of basis functions. 
We also compute the sum rule values for the quadrupole and monopole operators, and show that the DGCM yields more consistent results than the conventional GCM to the values from the double commutator. 
These results imply that the conjugate momentum is an important and relevant degree of freedom in collective motions.
\end{abstract}
\maketitle
\section{Introduction}
In strongly correlated quantum many-body systems such as nuclei, 
collective behaviors often emerge due to  
cooperation of many degrees of freedom.
These phenomena are called {\it collective motions} \cite{Ring_Schuck}, 
for which a translational motion, a rotational motion, a vibrational motion, and nuclear fission 
are typical examples. 
To describe them based on microscopic degrees of freedom 
is one of the major goals of many-body theory.

One of the basic microscopic theories describing quantum many-body systems is the mean-field theory.
In this theory, interacting quantum many-body systems are approximated by 
those of independent particles in a common mean-field. 
As particles move independently with each other, the many-body wave function in the mean-field approximation cannot include explicitly the quantum correlation among particles, except for the trivial Pauli correlation. 
Therefore, the pure mean-field theory is not designed to describe collective motions.

A way to resolve this problem is to mix several mean-field configurations. 
An important question in this method is what states should be mixed in the wave function.
The random phase approximation (RPA) can be regarded as a configuration mixing method, in which the configurations are restricted to one-particle one-hole (1p1h) excitations from a given mean field solution \cite{Ring_Schuck}.
Although the RPA is powerful in describing a small oscillation around the mean field solution, it is difficult to apply it to large amplitude motions, such as nuclear fission. 
In order to describe such large amplitude collective motions, the generator coordinate method (GCM) \cite{Ring_Schuck,HW53} has often been used in nuclear theory.
In this method, many-body states which are generated along some collective coordinate(s) are linearly superposed.
Many numerical calculations have been carried out in recent years \cite{Bender2003, NV11,E16,RR18,SO06,BH08,YM10,RT11,TR11,RE11,YB13,FS13,BA14,Yao2014,YZ15,EJ20}.

A serious theoretical ambiguity inherent in the GCM is that it is difficult to identify the appropriate collective coordinate for a motion of interest.
Notice that most of recent GCM calculations have been formulated based on a collective coordinate chosen empirically, but it is not obvious that this approach can appropriately describe the collective motion.
We mention that 
a collective coordinate can be unambiguously extracted in principle with 
the self-consistent collective coordinate method (SCC) \cite{MM80, M84} 
by projecting the collective motion onto a specific function space.
This is a self-contained theory, and does not require any empirical manipulation.
Furthermore, the collective coordinates and momenta are treated on the same footing.
However, 
the SCC is formulated in such a way that a mixing of configurations does not appear 
explicitly, and its application to the GCM is nontrivial. 
In this sense, the GCM is not yet a rigorous framework unlike the density functional theory (DFT) \cite{HK64, KS65} 
or the time-dependent DFT (TDDFT) \cite{RG84, NM06, CH12}, and one should carefully 
examine the validity of a chosen generated coordinate.
This problem has been recognized for a long time.

In the case of the translational motion, Peierls and Thouless pointed out that the naive GCM breaks certain symmetry, and 
they developed the double projection method \cite{PT62}. 
The double projection has also been formulated for the rotational motion.
Numerical calculations based on this method for the rotational motion are becoming possible with the development of computer technology \cite{BR15,EB16, R16, Shimada2016, ST15, Shimada2016, Ushitani2019}, 
even though it is still computationally rather expensive.

Since the symmetries are fundamental to this approach, the double projection has not been extended to more general collective motions.
In this connection, Goeke and Reinhard improved the GCM by introducing the conjugate momentum to the GCM.
This method is referred to as the dynamical GCM (DGCM) \cite{GR80}.
However, a practical application of this method has not been performed, except for the one with the Gaussian overlap approximation \cite{KE08}. 
In our previous work, we proposed the implementation of the DGCM based on the constrained variational method \cite{HH21}.
Based on this idea, we applied the DGCM to the particle number, which is numerically the easiest case.
This method, which could be regarded as the double projection on the number of particles, is also easy to implement due to the fact that the method of particle number projection is well known.
The purpose of this paper is to apply the DGCM actually to more general collective motions and to investigate the effect of the conjugate momentum on the collective motions.
For this purpose, we consider a quadrupole oscillation, which is often described with the conventional GCM.

The paper is organized as follows.
In Sec. II, we give a brief review of the GCM and the DGCM.
We also discuss how to perform the DGCM numerically based on the mean field theory for general collective motions.
In Sec. III, we carry out the DGCM calculation for $^{16}$O with a quadrupole operator and compare the results with those obtained with the conventional GCM. 
We then summarize the paper and discuss future perspectives in Sec. IV.

\section{Formulation}
\subsection{GCM}
In this section, we give a brief review of the GCM.
The GCM is a class of configuration mixing, which is performed along some generator coordinate $q$.
Therefore, the GCM starts from creating many-body wave functions $\{|q\rangle\}_q$ specified by the generator coordinate $q$. 
For a description of collective motions, $q$ is referred to as a collective coordinate. 
Of course, $q$ can be a complex number, and its dimension does not need to be specified.
However, for simplicity of the following discussion, we consider in this paper a one-dimensional real number for $q$. 
Then, the configuration mixing is expressed in the integral form as, 
\begin{equation}
    |\Psi\rangle_{\mathrm{GCM}}=\int dq\,f(q)|q\rangle,
\end{equation}
where $f(q)$ is called a weight function.
The subspace spanned by such a linear combination of $|q\rangle$ is called a collective subspace.
In the GCM, a Hamiltonian of a given system $\hat{H}$ is diagonalized by restricting to the collective subspace. 
Therefore, the fundamental equation of the GCM is obtained by projecting 
the time-independent Schr\"odinger equation onto the collective subspace,
\begin{equation}
    \langle q'|\left\{
    (\hat{H}-E)|\Psi \rangle_{\mathrm{GCM}}
    \right\} = 0, \quad \mathrm{for}\;\; {}^{\forall}|q'\rangle.
\end{equation}
We then obtain a generalized eigenvalue problem,
\begin{equation}
    \int dq\left(
    \langle q'|\hat{H}|q\rangle - E\langle q'|q\rangle
    \right)
    f(q) = 0,
\end{equation}
where $\mathcal{H}(q', q) := \langle q'|\hat{H}|q\rangle$ and $\mathcal{I}(q', q) := \langle q'|q\rangle$ are referred to as a Hamiltonian kernel and a norm kernel, respectively.
This equation is called the Hill-Wheeler equation \cite{HW53}.

The most serious problem in the GCM is how to prepare $|q\rangle$.
First, we have to decide what kind of function space to prepare the states $|q\rangle$.
In many cases, Slater determinant wave functions \cite{Bender2003, KK12} are used, but one may also superpose more complicated wave functions.
Next, we have to decide what method we use to draw the path $|q\rangle$ on these function spaces.
Above all things, a collective motion is not mathematically well defined in general, and thus the criteria of justification for a chosen $|q\rangle$ is unclear. 
The dynamical GCM, which we review in the next subsection, is theoretically more robust than the GCM and has certain good properties even when the collective operator $\hat{Q}$ is chosen empirically.

\subsection{Dynamical GCM}
The dynamical GCM introduced by Goeke and Reinhard can be regarded as the GCM that simultaneously considers the conjugate momentum $p$ in addition to a collective coordinate $q$ \cite{GR80}.
Therefore, as in the case of the GCM, the configuration mixing is performed as
\begin{equation}
  |\Psi\rangle_{\mathrm{DGCM}}=\iint dqdp\, f(q,p)|q,p\rangle,
  \label{dgcm}
\end{equation}
where this two-parameter path $|q, p\rangle$ is called a dynamical path.
The Hill-Wheeler equation is also obtained in the same way,
\begin{equation}
    \iint dqdp\left(
    \langle q',p'|\hat{H}|q,p\rangle - E\langle q',p'|q,p\rangle
    \right)
    f(q,p) = 0.
    \label{hwdgcm}
\end{equation}
Notice that Eqs. (\ref{dgcm}) and (\ref{hwdgcm}) alone do not guarantee that $p$ is the conjugate momentum of $q$, and one additional equation is required.
That is, the conjugate momentum of $p$ has to be determined to satisfy 
the conjugation condition,
\begin{equation}
    \langle q,p|
    \overleftarrow{\partial}_q\overrightarrow{\partial}_p
    -\overleftarrow{\partial}_p\overrightarrow{\partial}_q
    | q,p\rangle
    = i,
\end{equation}
where $\overleftarrow{\partial}_q$ means the partial derivative of $q$ acting on the left side and similar for $\overleftarrow{\partial}_p$. 
This condition can be seen as a kind of the canonical commutation relation.

Of course, this conjugate condition has to be supplemented by some boundary condition to determine the dynamical path. 
For example, we can construct the dynamical path $|q,p\rangle$ that satisfies the condition \cite{HH21},
\begin{equation}
   \label{eq:CC}
  |q,p\rangle = e^{ip\hat{Q}}|q\rangle,
\end{equation}
with the constrained condition, $\langle q|\hat{Q}|q\rangle = q$, 
where $\hat{Q}$ is some Hermitian operator related to a collective motion.
Since there are innumerable states satisfying the condition 
$\langle q|\hat{Q}|q\rangle = q$, the state $|q\rangle$ must be uniquely determined in some way.
When calculating $|q\rangle$ in the mean-field theory, one often uses a constrained variational method. 
Therefore, we shall use this method in this paper as well.
Considering the DGCM constructed in this way, one can show that under certain conditions the collective subspace defined with the DGCM has good properties \cite{hizawa22}.
As a result, an unexpected quantum entanglement is less likely to occur in the states obtained as solutions of the Hill-Wheeler equation. 
See Ref. \cite{hizawa22} for details.

It should be noted that the condition (\ref{eq:CC}) assumes the existence of the operator $\hat{Q}$.
Identifying $\hat{Q}$ for a given collective motion is a difficult problem in general and this remains so even with the DGCM. 
Therefore, we choose $\hat{Q}$ empirically in the rest of this paper.

\subsection{Dynamical path}
In performing numerical calculations with the DGCM, we need to consider specifically how to construct $e^{ip\hat{Q}}|q\rangle$.
We first make an assumption that $\hat{Q}$ is a one-body operator.
We then expand $\hat{Q}$ using some representation of creation and annihilation operators $\{\hat{c}^{\dagger}_i, \hat{c}_i\}_{i=1}^{\infty}$ as
\begin{equation}
  \hat{Q} = \sum_{i,j = 1}^{\infty}Q^{c}_{ij}\hat{c}^{\dagger}_i\hat{c}_j.
  \label{Q}
\end{equation}
Here, $Q^{c}_{ij}$ must be a Hermitian matrix because of Hermicity of $\hat{Q}$.
Using the Campbell-Baker-Hausdorff formula, we obtain
\begin{align}
  e^{ip\hat{Q}} \hat{c}^{\dagger}_ke^{-ip\hat{Q}} & = \sum_{l = 1}^{\infty}\left(e^{ipQ^c}\right)_{lk}\hat{c}^{\dagger}_l, \\
  e^{ip\hat{Q}} \hat{c}_ke^{-ip\hat{Q}} & = \sum_{l = 1}^{\infty}\left(e^{-ipQ^c}\right)_{lk}\hat{c}_l.
\end{align}
Let us consider the case where $|q\rangle$ is given in the Hartree-Fock approximation as, 
\begin{equation}
\label{eq:HF}
    |q\rangle = \prod_{i = 1}^{A}\hat{c}^{\dagger}_i|0\rangle,
\end{equation}
where $A$ is the number of nucleons and $|0\rangle$ is the vacuum state.
Note that we choose the representation of $\{\hat{c}^{\dagger}_i, \hat{c}_i\}_{i=1}^{\infty} $ so that $|q\rangle$ can be written as Eq. (\ref{eq:HF}).
In this case, the dynamical path is calculated as
\begin{equation}
  e^{ip\hat{Q}}|q\rangle
      = \prod_{i = 1}^{A}
      \left( \sum_{j_i} \left(e^{ipQ^c}\right)_{j_i i}
      \hat{c}^{\dagger}_{j_i}\right)|0\rangle.
      \label{dynamicalpath}
\end{equation}
Thus, the only thing which we have to do for preparing the dynamical path is to evaluate a unitary matrix,
\begin{equation}
  X^{\dagger}_{lk} := \left(e^{ipQ^c}\right)_{lk}.
\end{equation}
The largest advantage of using the Hartree-Fock approximation 
is that Eq. (\ref{dynamicalpath}) remains a Slater determinant since $X$ is a unitary matrix.
This implies that the conventional numerical technique of the GCM, i.e., the generalized Wick theorem, can be used also for the DGCM calculation. 
The same statement can be applied also to the Hartree-Fock-Bogoliubov approximation.

An extention to a more general case of $|q\rangle$ is trivial given that any state belonging to the Fock space is written by
\begin{equation}
  |\Psi\rangle = f(\{\hat{c}^{\dagger}_i, \hat{c}_i\}_{i=1}^{\infty})|0\rangle,
\end{equation}
where $f$ is any polynomial function.
We can calculate the dynamical path as in the case of the Hartree-Fock approximation,
\begin{equation}
  e^{ip\hat{Q}} |\Psi\rangle = f\left(\left\{\sum_{j}\hat{c}^{\dagger}_jX^{\dagger}_{ji},\,\, \sum_{j}X_{ij} \hat{c}_j\right\}_{i=1}^{\infty}\right)|0\rangle.
\end{equation}
In general, it is important to note that even when 
the original $|\Psi\rangle$ has a simple form it may not be so after the operator $e^{ip\hat{Q}}$ is multiplied to  $|\Psi\rangle$.
Therefore, even if we can evaluate $e^{ip\hat{Q}} |\Psi\rangle$, it is another issue whether the DGCM can be executed numerically.
In this regard, the scheme (\ref{eq:HF}) and (\ref{dynamicalpath}) based on the mean-field theory has the great advantage to implement the DGCM numerically.

In numerical cases, the full Fock space cannot be considered in practice.
Thus we must limit the Fock space to some model space.
We define the model space using a set of $M\in\mathbb{N}$ creation and annihilation operators, 
$\{\hat{b}^{\dagger}_i, \hat{b}_i\}_{i=1}^M$ and expand the operator $\hat{c}_i$ as 
\begin{equation}
  \hat{c}_i = \sum_{j = 1}^M C_{ij}\hat{b}_j,
\end{equation}
where $C$ is a $M\times M$ unitary matrix.
Then, we can determine the dynamical path using the operators
\begin{equation}
  \hat{d}_i = \sum_{j,k = 1}^{M}X_{ij}C_{jk}\hat{b}_k =: \sum_{j,k = 1}^{M}D_{ik}\hat{b}_k. 
\end{equation}
When we take the limit $M\to \infty$, the matrix $D$ becomes a unitary matrix. 
However, if $M$ is finite, $D$ is not a unitary matrix in general.
In such cases, we can not apply the generalized Wick theorem to the dynamical path, and the advantage of using the mean field theory is lost.

There are several approximate ways to resolve this problem. 
From a numerical point of view, the best way is to project the dynamical 
path onto the Slater determinant type wave functions. 
Such projected state is obtained by maximizing the overlap,
\begin{equation}
  \max_{|\psi\rangle}|\langle q, p|\psi\rangle|,
\end{equation}
where $|\psi\rangle$ spans the entire Slater determinant type wave functions inside the Fock space constructed by $\{\hat{b}^{\dagger}_i, \hat{b}_i\}_{i=1}^M$.
However, this method is computationally expensive and it is difficult to evaluate $X$ in many cases.
It would be easier to introduce a cutoff to Eq. (\ref{Q}) for the one body operator $\hat{Q}$ as
\begin{align}
  \hat{Q}^{\mathrm{cut}}
    & := \sum_{i,j = 1}^{M}Q^c_{ij}\hat{c}^{\dagger}_i\hat{c}_{j}
    = \sum_{i,j, k, l = 1}^{M}C^{\dagger}_{ki}Q^c_{ij}C_{jl}\hat{b}^{\dagger}_k\hat{b}_{l}. \\
    & =: \sum_{i,j = 1}^{M}Q^b_{ij}\hat{b}^{\dagger}_i\hat{b}_{j} \label{Qb}
\end{align}
Then, the dynamical path is calculated as
\begin{equation}
  X_{ij}^{\mathrm{cut}\,\dagger} = \left( e^{ipQ^c} \right)^{\mathrm{cut}}_{ij},
\end{equation}
where $(A)^{\mathrm{cut}}$ means that the matrix $A$ is evaluated as a $M \times M$ matrix. In this way, $X_{ij}^{\mathrm{cut}}$ is a $M \times M$ unitary matrix, and we can treat the dynamical path as a Slater determinant type wave function.

Using this simple cut-off method, we can calculate the dynamical path in the following steps.
At first, we diagonalize matrix $Q^b$ in Eq. (\ref{Qb}),
\begin{equation}
  Q^{b} = U\Lambda^{Q}U^{\dagger}, 
\end{equation}
where $\Lambda^{Q}$ is a diagonal matrix with the eigenvalues of $Q^b$. 
Note that all the matrices in this equation depend only on the basis representation $\{\hat{b}^{\dagger}_i, \hat{b}_i\}_{i=1}^{M}$ and the constraint operator $\hat{Q}$.
Next, we calculate the matrix $X^{\rm{cut}}$ as 
\begin{equation}
  \left(X^{\mathrm{cut}}\right)^{\dagger} = CUe^{ip\Lambda^Q}U^{\dagger}C^{\dagger}.
\end{equation}
Finally, we get
\begin{equation}
  D = CUe^{-ip\Lambda^Q}U^{\dagger}.
\end{equation}
In this way, the dynamical path is entirely determined by the matrix $D$ and one can only store $D$ in numerical calculations.

\section{Numerical applications of the DGCM to the quadrupole excitation}
\subsection{Numerical details}
In this paper, we apply the DGCM to the dynamical path specified by 
the quadrupole operator, which is often used in conventional GCM calculations.
This operator is defined as
\begin{equation}
    \hat{Q}_{20} =\int d^3r\,r^2Y_{20}(\theta, \phi)\sum_{\sigma, \tau}
    \hat{\psi}^{\dagger}(\bm{r}, \sigma, \tau) 
    \hat{\psi}(\bm{r}, \sigma, \tau) ,
\end{equation}
where $\hat{\psi}(\bm{r}, \sigma, \tau)$ is a nuclear field operator with spin $\sigma$ and isospin $\tau$.
Since this operator has the dimension of the square of the length, 
we introduce a dimensionless variable in a conventional way, 
\begin{equation}
    \hat{\beta} =\frac{\sqrt{20\pi}}{3r_0^2A^{5/3}}\hat{Q}_{20},
\end{equation}
where $r_0 = 1.2\,\mathrm{fm}$ and $A$ is the number of nucleons.
We use $\hat{\beta}$ for the collective operator $\hat{Q}$ in Eq. (\ref{Q}) in Sec. II.
In the following, we use the notation $(\beta, p_{\beta})$ for $(q,p)$.

We prepare the many-body state $|\beta\rangle$ using the constrained Hartree-Fock method, i.e.
\begin{equation}
    \delta\left\{
    \langle\hat{H}\rangle
    - \lambda(\langle\hat{\beta}\rangle - \beta)
    \right\} = 0, 
\end{equation} 
where the expectation value is defined as 
\begin{equation}
    {\langle\;\cdot\;\rangle} 
    := \frac{\langle\Psi|\;\cdot\;|\Psi\rangle}{\langle\Psi|\Psi\rangle},
\end{equation}
with the Slater determinant type wave function $|\Psi\rangle$. 
Here, $\lambda$ is a Lagrange multiplier.
Solving this equation, we obtain the normalized DGCM (GCM) basis function $|\beta\rangle$ satisfying $\langle\beta|\hat{\beta}|\beta \rangle = \beta$.
For the Hamiltonian $\hat{H}$, we use the Gogny interaction with the D1S 
parameter set \cite{DG80, BG91}\footnote{Note that the sign of the spin-orbit force parameter is wrong in the original D1S paper \cite{BG91}}.
The Coulomb interaction is treated without any approximation, and the center-of-mass correction is evaluated by subtracting the energy of the center-of-mass motion from $\hat{H}$ beforehand.
Therefore, the Hamiltonian $\hat{H}$ used in this paper reads, 
\begin{equation}
    \hat{H} = \sum_{i=1}^{A}\frac{\hat{\bm{p}}_i^2}{2m}
    +\frac{1}{2}\sum_{i\neq j}\hat{V}(i, j)
    -\frac{1}{2mA}\left(\sum_{i=1}^{A}\hat{\bm{p}}_i\right)^2,
\end{equation}
where $\hat{\bm{p}}_i$ is the momentum operator of the $i$-th particle, 
and $\hat{V}(i, j)$ means a sum of the Gogny and the Coulomb interactions between the $i$-th and the $j$-th 
particles.
Since both the Gogny and the Coulomb interactions are finite range interactions, the basis expansion method is effective.
To this end, we use the 3-dimensional spherical harmonic oscillator basis,
\begin{align}
    \phi^{\mathrm{HO}}_{n_x, n_y, n_z}(\bm{r}) =&
    H_{n_x}\left(\frac{x}{b}\right)
    H_{n_y}\left(\frac{y}{b}\right)
    H_{n_z}\left(\frac{z}{b}\right) \notag \\
    &\times \exp\left(-\frac{r^2}{2b^2}\right)
    \times \mathrm{normalization},
\end{align}
where $b$ is the harmonic oscillator length and $H_{n}$ is a Hermitian polynomial with the quantum number $n\in\mathbb{N}\cup\{0\}$.
The basis functions are truncated at $N=n_x+n_y + n_z\leq N_0$, which leads to $(N_0+1)(N_0+2)(N_0+3)/6$ bases. 
The spin wave function is introduced in the form of a direct product with the harmonic oscillator basis, while the isospin mixing is not considered. 
Taking into account the spin degree of freedom, the total number of bases is $(N_0+1)(N_0+2)(N_0+3)/3$ for protons and neutrons, respectively.
Note that we use the Hartree-Fock approximation in this paper, ignoring the pairing effect.

The dynamical path $|\beta, p_{\beta}\rangle= e^{i p_{\beta}\hat{\beta}}|\beta\rangle$ is prepared according to Sec. II.
Notice that $|\beta, 0\rangle$ so constructed is equal to $|\beta\rangle$.
In this case, the cut-off dimension $M$ in Sec. II C is the same as $(N_0+1)(N_0+2)(N_0+3)/3$ for neutrons and protons, respectively. 
We evaluate the Hamiltonian kernel in the DGCM (GCM) using the generalized Wick theorem \cite{BB69}.
Notice that the Gogny D1S interaction is the effective one and has the density dependent term proportional to the 1/3 power of the nucleon number density.
Since the generalized Wick theorem cannot be treated rigorously in this class of effective interaction, we use an ansatz of substituting the local transition density in the density-dependent part \cite{BD90}.

Finally, we mention the issue of overcompleteness.
In general, $\{|\beta\rangle\}_{\beta}$ or $\{|\beta, p_{\beta}\rangle\}_{\beta, p_{\beta}}$ are not linearly independent to each other.
This linear dependence implies that the Hill-Wheeler equation has solutions with indefinite energy.
This is referred to as the overcompleteness problem.
We can resolve it by artificially removing the zero eigenvalue modes of the norm kernel.
However, in numerical calculations, the problem of overcompleteness becomes more serious.
This is because small eigenvalues of the norm kernel may cause numerical problems due to numerical errors, even when the eigenvalues are not exactly zero. 
Therefore, we must remove the small eigenvalue modes by hand to achieve stable numerical calculations.
In this paper, we remove the those modes with eigenvalues less than $10^{-5}$.

\subsection{Results}
We now apply the GCM and the DGCM to the quadrupole excitations of $^{16}$O.  
To this end, we include the harmononic oscillator basis up to $N_0=10$ with the oscillator length of $b = 1.457433\;\mathrm{fm}$, which is the optimized value in the mean-field approximation with HFBTHO \cite{HFBTHOver3.0}. 
We first generate the coordinate $\beta$ in increments of 0.02 in the range of $-$0.24 to 0.24.
The range of $\beta$ is chosen so that a level crossing does not occur, as we do not include the pairing correlation. 
We then generate the conjugate momentum $p_{\beta}/2\pi$ from $-$0.6 to 0.6 in increments of 0.05 for each $\beta$.

\begin{figure} [btp]
    \includegraphics[scale=0.68,clip]{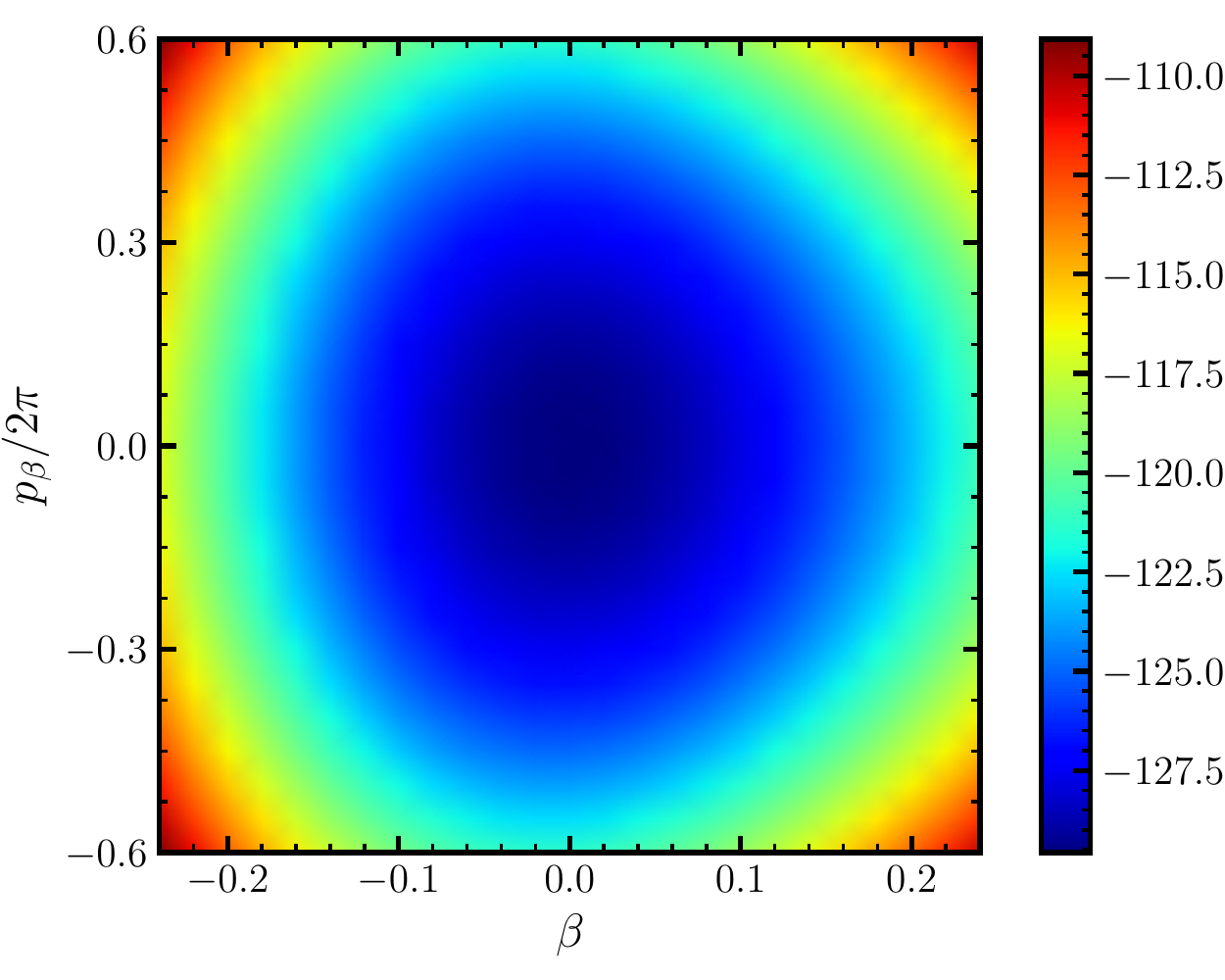}
    \caption{The energy surface of the dynamical path $E(\beta, p_{\beta})$ for the quadrupole motion of $^{16}$O, in units of MeV. 
    The Gogny D1S interaction is used to draw this figure. 
    The minimum of the energy surface is $E(0, 0) = -129.569\,\mathrm{MeV}$, which is the same as the Hartree-Fock result without the constraint condition.
    Because of the time reversal symmetry of $|\beta, 0\rangle$,  the energy surface is symmetric about the $p_{\beta}-$axis.
    }
    \label{fig:ES}
\end{figure}

Figure \ref{fig:ES} shows the energy surface defined as $E(\beta, p_{\beta}):=\langle \beta, p_{\beta}|\hat{H}|\beta, p_{\beta}\rangle$. 
As we can see, we determine the range of $p_{\beta}$ from the condition that $E(\pm0.24, 0)$ and $E(0, \pm0.6 \times 2\pi)$ are similar to each 
other. 
Since $|\beta\rangle = |\beta, 0\rangle$ is an eigenstate of the time reversal operator in our method, $|\beta, p_\beta\rangle$ and $|\beta, -p_\beta\rangle$ are paired for the time reversal operation.
This is leads to the inverted symmetry of the energy surface with respect to the vertical direction in Fig. \ref{fig:ES}.

\begin{figure} [btp]
    \includegraphics[scale=0.68,clip]{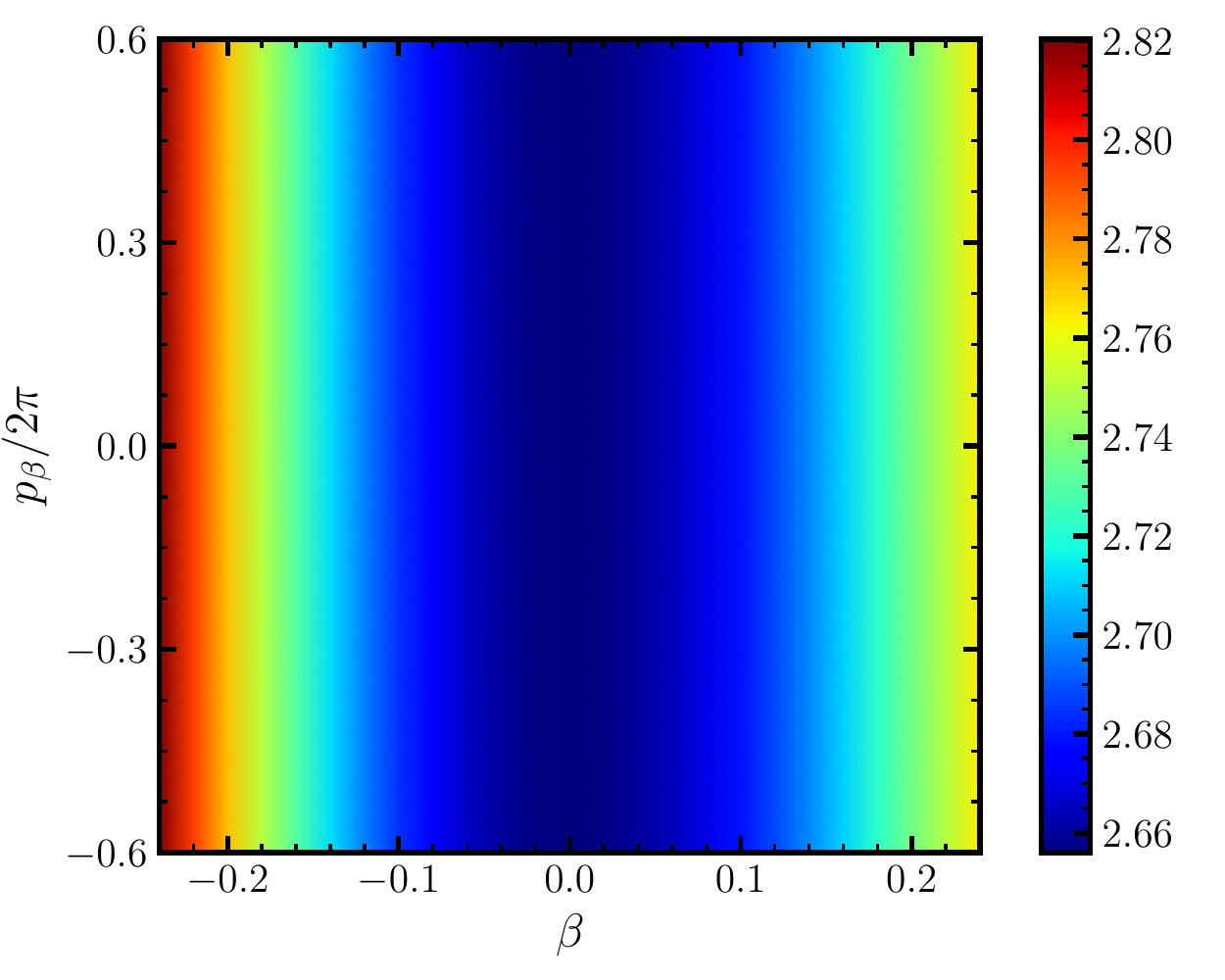}
    \caption{The root-mean-square (rms) radius for the dynamical path $r_{\mathrm{rms}}(\beta, p_{\beta})$, in units of fm.
    The rms radius increases as the deformation $\beta$ increases, while it is flat in the direction of $p_{\beta}$. 
    }
    \label{fig:RS}
\end{figure}

Figure \ref{fig:RS} shows the root-mean-square (rms) radius $r_{\mathrm{ms}}(\beta, p_{\beta})$ for each basis, which is defined as
\begin{equation}
    r_{\mathrm{rms}}(\beta, p_{\beta})
    = \sqrt{\langle \beta, p_{\beta}|
    \int d^3r\,r^2\hat{\rho}(\bm{r})|\beta, p_{\beta}\rangle / A},
\end{equation}
where $\hat{\rho}(\bm{r})$ is the particle number density operator of nucleons.
Notice that $r_{\mathrm{rms}}(\beta, p_{\beta})$ is almost independent of $p_{\beta}$. 
This means that the operator $e^{ip_{\beta}\hat{\beta}}$ does not significantly change the rms radius.
This result is not surprising, given that $\hat{\beta}$ and $\int d^3r\,r^2\hat{\rho}(\bm{r})$ are commutative in the full space.
However, rigorously speaking, they are not commutative, since we now restrict the model space to a finite space.
Fig. \ref{fig:RS} implies that our model space is large enough and this noncommutativity is practically 
negligible.
Notice that, as in the rms radius, the operator $e^{ip_{\beta}\hat{\beta}}$ in $|\beta,p_\beta\rangle$ keeps the expectation value of any local deformation operator,
\begin{equation}
    \hat{Q} = \sum_{\sigma, \tau, \sigma', \tau'}\int d^3r \, Q(\bm{r}, \sigma, \tau, \sigma', \tau')
    \hat{\psi}^{\dagger}(\bm{r}, \sigma, \tau) 
    \hat{\psi}(\bm{r}, \sigma', \tau'), 
\end{equation}
invariant in the full space.  
This implies that dealing with the dynamical path in the DGCM allows us to introduce the effect of internal excitations for each $\langle\hat{Q}\rangle$.
That is, $|\beta, p_\beta\rangle$ all have the same expectation value of $\hat{Q}$ for each $\beta$, while the energy expectation value is different.
This effect cannot be captured by the conventional GCM with $|\beta\rangle$.

In performing the DGCM, it is computationally expensive to include all of the 625 points in the $(\beta,p_\beta)$ plane. 
We thus select 25 points, $S^{\mathrm{DGCM}}_{25} = \{(\beta, p_{\beta})\,|\, \beta \in \{-0.2, -0.1, 0, 0.1, 0.2\}, \quad p_{\beta} /2\pi \in \{-0.6, -0.3, 0, 0.3, 0.6\}\}$, to carry out the DGCM calculations. 
On the other hand, for the GCM calculations, we use 25 points of 
$S^{\mathrm{GCM}}_{25} = \{(\beta, 0)\,|\, \beta \in \{-0.24, -0.22, \cdots, 0.24\}\}$, to compare the DGCM to the GCM under the same conditions as much as possible.
\begin{table}[h]
 \caption{The GCM and the DGCM energy for the quadrupole excitations of $^{16}$O with the point sets $S^{\mathrm{GCM}}_{25}$ and $S^{\mathrm{DGCM}}_{25}$, respectively. 
 The cut-off for the norm kernel is taken as $10^{-5}$.
  }
 \centering
 \label{tab:E}
  \begin{tabular}{c|cc}
   \hline\hline
    state   &   GCM (MeV)   &   DGCM (MeV)   \\
   \hline 
   1  & $-$129.682 & $-$129.765 \\
   2  & $-$107.993 & $-$108.140 \\
   3  & $-$92.260 & $-$104.475 \\
   4  & $-$77.911 & $-$87.019 \\
   5  & $-$64.097 & $-$83.059 \\
   \hline\hline
  \end{tabular}
\end{table}
TABLE \ref{tab:E} shows the results of the GCM and the DGCM energies with these point sets, obtained with 
the cut-off for the norm kernel of $10^{-5}$.
We can see that the energies are lower for the DGCM for all the states shown in the table, even though we use the same number of basis functions.
This tendency does not change even when the cutoff is decreased to $10^{-10}$.

The better performance of the DGCM can be understood by investigating 
the eigenvalue distributions of the norm kernel. 
In the DGCM, the smallest eigenvalue is O($10^{-8}$).
On the other hand, in the GCM, there are 10 eigenvalues which are less than 10$^{-13}$. 
This implies that the DGCM contains many more relevant states that are orthogonal to each other.

\begin{figure*} [tbp]
    \includegraphics[scale=0.7,clip]{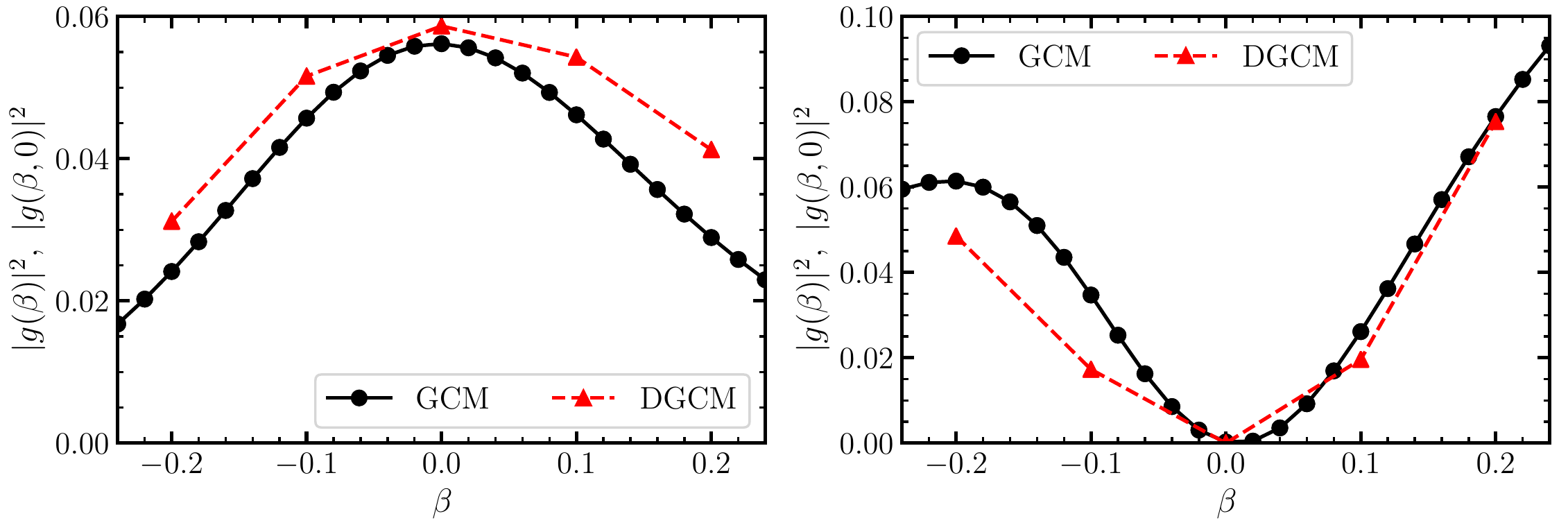}
    \caption{The square of the collective wave functions for the ground (the left panel) and the first excited (the right panel) 
    states. The black solid and the red dashed lines represent the collective wave functions 
    for the GCM and the DGCM, respectively. For the DGCM wave function, $p_{\beta}$ is set to be 0.
    }
    \label{fig:wf_GCM}
\end{figure*}

\begin{figure*} [tbp]
    \includegraphics[scale=0.68,clip]{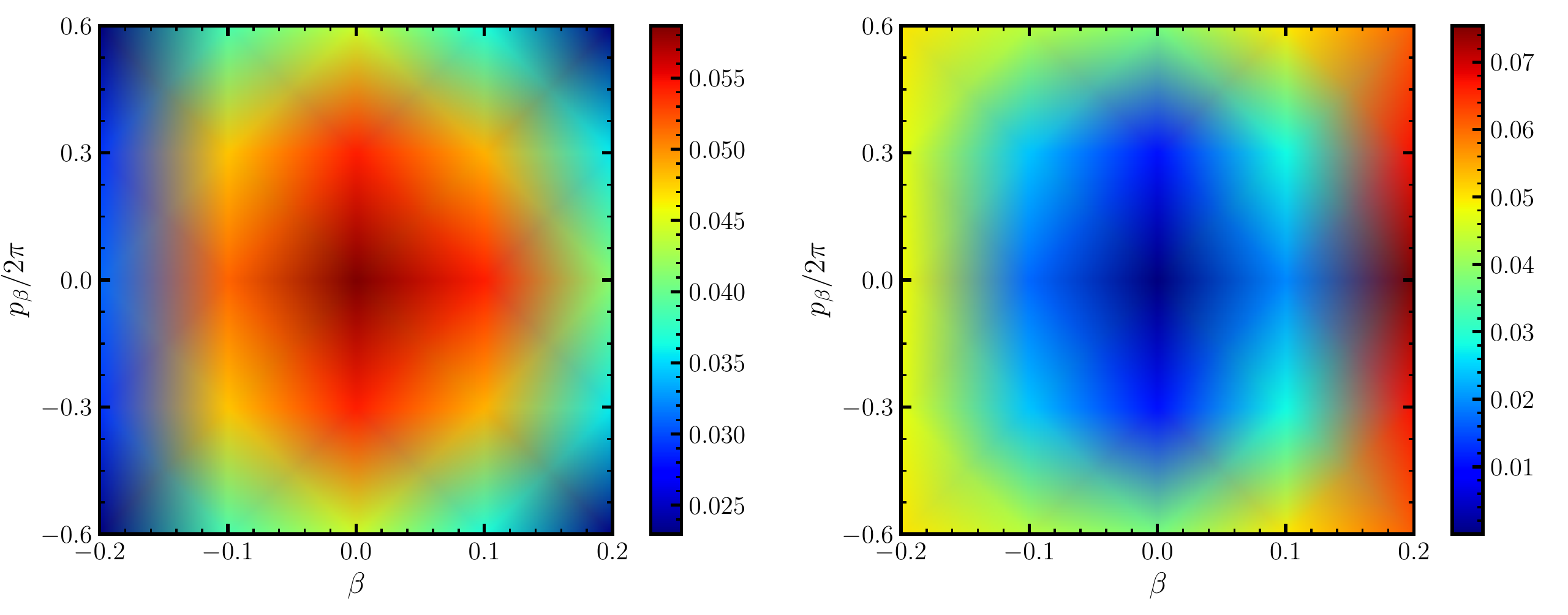}
    \caption{The square of the collective wave functions for the ground (the left panel) and the first excited (the right panel) 
    states of the DGCM in the two-dimensional $(\beta,p_\beta)$ plane. 
    }
    \label{fig:wf_DGCM}
\end{figure*}

Figures \ref{fig:wf_GCM} and \ref{fig:wf_DGCM} show the square of the GCM collective wave function $|g(\beta)|^2$ and that of the DGCM $|g(\beta, p_{\beta})|^2$.
The left and the right panels in each figure show the collective wave functions for the ground and the first excited states, respectively.
In the GCM case, the collective wave function is defined as
\begin{equation}
    g(\beta)=\int d\beta' \,\mathcal{I}^{1/2}(\beta, \beta')f(\beta'),
\end{equation}
where $\mathcal{I}^{1/2}$ is the square root of the norm kernel.
In this case, the collective wave functions for different GCM eigenstates 
are orthogonal to each other.
The corresponding collective wave function for the DGCM, $g(\beta, p_{\beta})$, is defined in a similar way.
We can see that the collective wave functions are harmonic oscillator-like in both the GCM and the DGCM cases.
As is evidenced in the collective wave functions of the DGCM, which are spread over both $\beta$ and $p_{\beta}$, the conjugate momentum is an important degree of freedom.

Finally, we discuss the results of the sum rule.
The sum rule in the GCM is already known and can be formulated as in the usual case \cite{FV76}.
See Eqs. (\ref{eq:SR_mono}) and (\ref{eq:SR_Q20}) in Appendix A for the explicit forms of the sum rules for the monopole operator, $\hat{Q}_0 = \sum_{i=1}^{A}\hat{\bm{r}}_i^2$ and the quadrupole operator $\hat{Q}_{20} = \sum_{i=1}^{A}(2\hat{z}_i^2 -\hat{x}_i^2 - \hat{y}_i^2)$, respectively. 
Note that the validity of the usual sum rule depends on the structure of the subspace (see Appendix A).
This fact is known numerically \cite{JL20}, and it can be seen also with the DGCM and the GCM.
TABLE \ref{tab:SR} shows the sum rule results for the quadrupole and the monopole operators based on the GCM and the DGCM.
\begin{table}[h]
 \caption{
 Results of the sum rule calculations for the quadrupole and the monopole operators.
 The values in this table show the ratios of the left-hand side to the right-hand side of Eqs. (\ref{eq:SR_mono}) and (\ref{eq:SR_Q20}).
 }
 \centering
 \label{tab:SR}
  \begin{tabular}{c|cc|cc}
   \hline \hline
   state & \multicolumn{2}{c|}{quadrupole} & \multicolumn{2}{c}{monopole} \\
   \hline
       $i$  & GCM & DGCM  & GCM & DGCM  \\
   \hline 
   1  &  1.0005 & 1.0047 & 0.3855 & 0.9176 \\
   2  &  0.6287 & 1.0162 & 0.6533 & 0.9926 \\
   3  &  0.7309 & 1.0512 & 0.6184 & $-$0.0066 \\
   4  &  0.6370 & 0.9414 & 0.5818 & 0.6449 \\
   5  & $-$0.5878 & 0.6352 &$-$0.8390 & $-$0.0981 \\
   \hline \hline
  \end{tabular}
\end{table}
The values in this table are the ratios of the left-hand side to the right-hand side of Eqs. (\ref{eq:SR_mono}) and (\ref{eq:SR_Q20}), which should ideally be 1 if the sum rules are perfectly satisfied.
The label $i$ denotes the DGCM (or the GCM) states corresponding to those in Table \ref{tab:E}.
For the quadrupole operator, both the GCM and the DGCM are close to unity for the ground state ($i = 1$).
In the DGCM, this is the case also for the excited states up to the third excited state ($i=4$).
In contrast, the sum rule is satisfied much less satisfactory in the GCM. 
It can thus be concluded that the DGCM is more likely to include relevant states than the GCM for the quadrupole oscillation.
This tendency does not change even when we use a smaller cut-off for the eigenvalues of the norm kernel.
For example, if we take the cut-off of $10^{-10}$, the sum rule for the excited states remain unsatisfactory in the GCM, 
whereas that in the DGCM is good up to the 7-th excited state.
This is related to the fact that under certain ansatzs the collective subspace in the DGCM can be represented as a tensor product of the collective and the non-collective degrees of freedom \cite{hizawa22}.

Similarly, the results for the monopole operator indicate that the DGCM is more superior than the GCM.
Therefore, the DGCM can prepare different states more efficiently than the GCM. 
However, we find that smaller cutoffs do not improve the results much, both for the GCM and the DGCM.
This fact indicates that the collective subspace in this calculation does not contain enough degrees of freedom relevant to the monopole operator. 
This is a natural consequence of the fact that the collective subspace is prepared in this calculations primarily for the quadrupole motion, without caring about the monopole motion.

\section{Summary and future perspectives}
In this paper, we have discussed how to perform actual numerical calculations for the DGCM, which is an extension of the GCM by including the conjugate momentum.
We have provided a concrete scheme to construct the dynamical path approximately for general one-body operators.
Using this scheme, we have applied the DGCM to the quadrupole oscillation of ${}^{16}\mathrm{O}$ and have compared the results with those of the conventional GCM.
We have found that the DGCM leads to more consistent results than the GCM both for the energies and the sum rules. 
This suggests that, at least for the quadrupole oscillations, it is essential to take into account the conjugate momentum.

In this paper, we have neglected the pairing correlation and used the Hartree-Fock approximation. 
An obvious future extention of our work is to use the Hartree-Fock-Bogoliubov (HFB) approximation, with which the pairing effect can be fully considered.
Such extension will be important to discuss large amplitude collective motions, especially a nuclear fission. 
Since the HFB breaks the U(1) symmetry with respect to the particle number, 
we should perform a particle number projection to restore the symmetry.
The methods in this paper can be applied immediately to this case.
It will be interesting to see how the conjugate momentum alters the results of the conventional calculation of the nuclear fission with the WKB approximation.

Theoretical aspects of the DGCM themselves are also interesting topics to explore. 
Many conventional GCM calculations are based on microscopic degrees of freedom, but have a strong phenomenological factor.
The main reason for this is that one does not have good information on a collective subspace.
It is expected that the DGCM is relatively easy to grasp the structure of the space.
This is one of the largest advantages of the DGCM, which allows us to discuss new aspects for the collective motion, such as entanglement entropy.

\section*{Acknowledgments}
The authors thank Y. Hashimoto for numerical advice and N. Shimizu for useful discussions.
This work was supported by JSPS KAKENHI
(Grant Nos. 21J22348, JP19K03824, JP19K03861, and JP19K03872).

\appendix

\section{Sum Rules in a truncated space}
\subsection{A general consideration}
Let us consider a Hilbert space $\mathcal{H}_{\mathrm{tot}}$ and introduce 
linear operators $\hat{A}$ and $\hat{B}$ which map from $\mathcal{H}_{\mathrm{tot}}$ to itself, $\mathcal{H}_{\mathrm{tot}}$, that is, $\mathcal{H}_{\mathrm{tot}}\to \mathcal{H}_{\mathrm{tot}}$.
We assume that $\hat{A}$ and $\hat{B}$ are Hermitian operators.
We then consider an orthonormal subset $\{|i\rangle\}_{i\in\mathcal{X}}\subset \mathcal{H}_{\mathrm{tot}}$ that satisfies 
\begin{equation}
    \label{eq:A}
    \langle i|\hat{A}| j\rangle
    = a_i\langle i| j\rangle
    = a_i\delta_{ij}, \quad i,j\in \mathcal{X}.
\end{equation}
Because $\hat{A}$ is a Hermite operator, $a_i$ is a real number.
Note that this equation does not necessarily mean that $|i\rangle$ is an eigenstate of $\hat{A}$ in the full space.
That is, because of a truncation, the subspace $\mathcal{H}_{\mathrm{sub}}$ spanned by $\{|i\rangle\}_{i\in\mathcal{X}}$ is smaller than $\mathcal{H}_{\mathrm{tot}}$, $\mathcal{H}_{\mathrm{sub}}\subseteq\mathcal{H}_{\mathrm{tot}}$.
Equation (\ref{eq:A}) is simply interpreted as a diagonalization of $\hat{A}$ within a restricted space. 
The projection operator onto $\mathcal{H}_{\mathrm{sub}}$ can then be constructed as
\begin{equation}
    \hat{\pi}_{\mathrm{sub}}=\sum_{i\in\mathcal{X}}|i\rangle\langle i|.
\end{equation}
This is an identity operator of $\mathcal{H}_{\mathrm{sub}}$.

We consider the following variable,
\begin{gather}
    S_{AB}(i) :=\sum_{j\in\mathcal{X}}(a_j-a_{i})|\langle j|\hat{B}|i\rangle|^2
\end{gather}
When $\hat{A}$ is the Hamiltonian $\hat{H}$, this is nothing but the energy weighted sum rule. 
If we assume $[\hat{B}, \hat{\pi}_{\mathrm{sub}}]=0$, we find 
\begin{align}
    S_{AB}(i)  & =
    \sum_{j\in\mathcal{X}}
    (\langle j|a_j \hat{B}|i\rangle
    - \langle j| \hat{B}a_{i}|i\rangle)
    \langle i| \hat{B}|j\rangle \\
    & = 
    \langle i| \hat{B}[\hat{A}, \hat{B}]|i\rangle.
\end{align}
Since $S_{AB}(i)$ is real, this can also be written as 
\begin{align}
    S_{AB}(i) &
    = \frac{1}{2}(S_{AB}(i) + S_{AB}(i)^*) \\
    & =
    \label{eq:SRG}
    \frac{1}{2}\langle i|[\hat{B}, [\hat{A}, \hat{B}]]|i\rangle.
\end{align}
Note that Eq. (\ref{eq:SRG}) is trivially satisfied under the assumption $[\hat{A}, \hat{\pi}_{\mathrm{sub}}]=0$.
Of course, in this case, $[\hat{B}, \hat{\pi}_{\mathrm{sub}}]$ does not  have to be zero.

\subsection{Explicit form for the quadrupole and the monopole fields}
We here consider the case in which the operator $\hat{A}$ is a Hamiltonian $\hat{H}$ and the operator $\hat{B}$ is a local one-body operatyor $\sum_{i\in \mathcal{Y}}f(\hat{\bm{r}}_i)$ with some $\mathcal{Y}\subseteq\{1, 2, \cdots, A\}$.
Since the Hamiltonian $\hat{H}$ usually contains only up to the second order of the single particle momentum operator $\hat{\bm{p}}_{\alpha}$, we can evaluate Eq. (\ref{eq:SRG}) explicitly.
Notice that the canonical commutation relation leads to a useful formula,
\begin{equation}
    [\hat{\bm{p}}_{\alpha}, f(\hat{\bm{r}}_i)] =-i\hbar \nabla f(\hat{\bm{r}}_i)\delta_{i,\alpha}.
\end{equation}
We then obtain
\begin{align}
    [f(\hat{\bm{r}}_i), 
    [\hat{\bm{p}}_{\alpha}\cdot \hat{\bm{p}}_{\beta}
    , f(\hat{\bm{r}}_k)]]
    & =
    \hbar^2
    \nabla f(\hat{\bm{r}}_i)\cdot \nabla f(\hat{\bm{r}}_k) 
    \delta_{i,\alpha}\delta_{k,\beta}
    \notag \\
    & +
    \hbar^2
    \nabla f(\hat{\bm{r}}_k)\cdot \nabla f(\hat{\bm{r}}_i) 
    \delta_{i,\beta}\delta_{k,\alpha}.
\end{align}
Using these equations, one finally obtains, 
\begin{align}
    [\hat{B}, [\hat{H}, \hat{B}]]
    & =
    \frac{\hbar^2}{m}\left(
    1-\frac{1}{A}
    \right)
    \sum_{i\in\mathcal{Y}}(\nabla f(\hat{\bm{r}}_i))^2
    \notag \\
    & -
    \frac{\hbar^2}{mA}
    \sum_{\substack{i,k\in\mathcal{Y} \\ i \neq k}}\nabla f(\hat{\bm{r}}_k)\cdot \nabla f(\hat{\bm{r}}_i).
\end{align}
The first term in this equation is due to the contributions of the kinetic energy term and the one-body part of the center-of-mas correction, while the two-body part of the center-of-mass correction yields the second term.

In this paper, we consider the monopole operator $\hat{Q}_0 = \sum_{i=1}^{A}\hat{\bm{r}}_i^2$ and the quadrupole operator $\hat{Q}_{20} = \sum_{i=1}^{A}(2\hat{z}_i^2 -\hat{x}_i^2 - \hat{y}_i^2)$ for concrete examples of $\hat{B}$.
Note that their normalization factors need not be taken into account.
Therefore, we obtain the sum rules
\begin{gather}
    \label{eq:SR_mono}
    S_{H, Q_0}(i) 
     =\frac{2\hbar^2}{m}\left(1-\frac{1}{A}\right)\langle i|\hat{Q}_0|i\rangle
    -\frac{2\hbar^2}{mA}\langle i|
    \sum_{\substack{i,k=1 \\ i \neq k}}^{A}\hat{\bm{r}}_i \cdot \hat{\bm{r}}_k
    |i\rangle, \\
    \label{eq:SR_Q20}
    S_{H, Q_{20}}(i) 
     =\frac{2\hbar^2}{m}\left(1-\frac{1}{A}\right)\langle i|\hat{Q}_{a}|i\rangle
    -\frac{2\hbar^2}{mA}\langle i|
    \hat{Q}_b
    |i\rangle, 
\end{gather}
where the operators $\hat{Q}_a$ and $\hat{Q}_b$ are defined as 
\begin{equation}
    \hat{Q}_{a} = \sum_{i = 1}^{A}
    (4\hat{z}_i^2 + \hat{x}_i^2 + \hat{y}_i^2), 
\end{equation}
    and 
\begin{equation}
    \hat{Q}_{b} = \sum_{\substack{i,k=1 \\ i \neq k}}^{A}
    (4\hat{z}_i \hat{z}_k + \hat{x}_i \hat{x}_k + \hat{y}_i \hat{y}_k),
\end{equation}
respectively.

\bibliographystyle{apsrev4-2}
\bibliography{ref}

\begin{thebibliography}{44}%
\makeatletter
\providecommand \@ifxundefined [1]{%
 \@ifx{#1\undefined}
}%
\providecommand \@ifnum [1]{%
 \ifnum #1\expandafter \@firstoftwo
 \else \expandafter \@secondoftwo
 \fi
}%
\providecommand \@ifx [1]{%
 \ifx #1\expandafter \@firstoftwo
 \else \expandafter \@secondoftwo
 \fi
}%
\providecommand \natexlab [1]{#1}%
\providecommand \enquote  [1]{``#1''}%
\providecommand \bibnamefont  [1]{#1}%
\providecommand \bibfnamefont [1]{#1}%
\providecommand \citenamefont [1]{#1}%
\providecommand \href@noop [0]{\@secondoftwo}%
\providecommand \href [0]{\begingroup \@sanitize@url \@href}%
\providecommand \@href[1]{\@@startlink{#1}\@@href}%
\providecommand \@@href[1]{\endgroup#1\@@endlink}%
\providecommand \@sanitize@url [0]{\catcode `\\12\catcode `\$12\catcode
  `\&12\catcode `\#12\catcode `\^12\catcode `\_12\catcode `\%12\relax}%
\providecommand \@@startlink[1]{}%
\providecommand \@@endlink[0]{}%
\providecommand \url  [0]{\begingroup\@sanitize@url \@url }%
\providecommand \@url [1]{\endgroup\@href {#1}{\urlprefix }}%
\providecommand \urlprefix  [0]{URL }%
\providecommand \Eprint [0]{\href }%
\providecommand \doibase [0]{https://doi.org/}%
\providecommand \selectlanguage [0]{\@gobble}%
\providecommand \bibinfo  [0]{\@secondoftwo}%
\providecommand \bibfield  [0]{\@secondoftwo}%
\providecommand \translation [1]{[#1]}%
\providecommand \BibitemOpen [0]{}%
\providecommand \bibitemStop [0]{}%
\providecommand \bibitemNoStop [0]{.\EOS\space}%
\providecommand \EOS [0]{\spacefactor3000\relax}%
\providecommand \BibitemShut  [1]{\csname bibitem#1\endcsname}%
\let\auto@bib@innerbib\@empty
%</preamble>
\bibitem [{\citenamefont {Ring}\ and\ \citenamefont
  {Schuck}(1980)}]{Ring_Schuck}%
  \BibitemOpen
  \bibfield  {author} {\bibinfo {author} {\bibfnamefont {P.}~\bibnamefont
  {Ring}}\ and\ \bibinfo {author} {\bibfnamefont {P.}~\bibnamefont {Schuck}},\
  }\href {https://link.springer.com/book/9783540212065} {\emph {\bibinfo
  {title} {The nuclear many-body problem}}}\ (\bibinfo  {publisher}
  {Springer-Verlag},\ \bibinfo {address} {New York},\ \bibinfo {year}
  {1980})\BibitemShut {NoStop}%
\bibitem [{\citenamefont {Hill}\ and\ \citenamefont {Wheeler}(1953)}]{HW53}%
  \BibitemOpen
  \bibfield  {author} {\bibinfo {author} {\bibfnamefont {D.~L.}\ \bibnamefont
  {Hill}}\ and\ \bibinfo {author} {\bibfnamefont {J.~A.}\ \bibnamefont
  {Wheeler}},\ }\href {https://doi.org/10.1103/PhysRev.89.1102} {\bibfield
  {journal} {\bibinfo  {journal} {Phys. Rev.}\ }\textbf {\bibinfo {volume}
  {89}},\ \bibinfo {pages} {1102} (\bibinfo {year} {1953})}\BibitemShut
  {NoStop}%
\bibitem [{\citenamefont {Bender}\ \emph {et~al.}(2003)\citenamefont {Bender},
  \citenamefont {Heenen},\ and\ \citenamefont {Reinhard}}]{Bender2003}%
  \BibitemOpen
  \bibfield  {author} {\bibinfo {author} {\bibfnamefont {M.}~\bibnamefont
  {Bender}}, \bibinfo {author} {\bibfnamefont {P.-H.}\ \bibnamefont {Heenen}},\
  and\ \bibinfo {author} {\bibfnamefont {P.-G.}\ \bibnamefont {Reinhard}},\
  }\href {https://doi.org/10.1103/RevModPhys.75.121} {\bibfield  {journal}
  {\bibinfo  {journal} {Rev. Mod. Phys.}\ }\textbf {\bibinfo {volume} {75}},\
  \bibinfo {pages} {121} (\bibinfo {year} {2003})}\BibitemShut {NoStop}%
\bibitem [{\citenamefont {Nikšić}\ \emph {et~al.}(2011)\citenamefont
  {Nikšić}, \citenamefont {Vretenar},\ and\ \citenamefont {Ring}}]{NV11}%
  \BibitemOpen
  \bibfield  {author} {\bibinfo {author} {\bibfnamefont {T.}~\bibnamefont
  {Nikšić}}, \bibinfo {author} {\bibfnamefont {D.}~\bibnamefont {Vretenar}},\
  and\ \bibinfo {author} {\bibfnamefont {P.}~\bibnamefont {Ring}},\ }\href
  {https://doi.org/https://doi.org/10.1016/j.ppnp.2011.01.055} {\bibfield
  {journal} {\bibinfo  {journal} {Prog. Part. Nucl. Phys.}\ }\textbf {\bibinfo
  {volume} {66}},\ \bibinfo {pages} {519 } (\bibinfo {year}
  {2011})}\BibitemShut {NoStop}%
\bibitem [{\citenamefont {Egido}(2016)}]{E16}%
  \BibitemOpen
  \bibfield  {author} {\bibinfo {author} {\bibfnamefont {J.~L.}\ \bibnamefont
  {Egido}},\ }\href {https://doi.org/10.1088/0031-8949/91/7/073003} {\bibfield
  {journal} {\bibinfo  {journal} {Phys. Scr.}\ }\textbf {\bibinfo {volume}
  {91}},\ \bibinfo {pages} {073003} (\bibinfo {year} {2016})}\BibitemShut
  {NoStop}%
\bibitem [{\citenamefont {Robledo}\ \emph {et~al.}(2018)\citenamefont
  {Robledo}, \citenamefont {Rodr{\'{\i}}guez},\ and\ \citenamefont
  {Rodr{\'{\i}}guez-Guzm{\'{a}}n}}]{RR18}%
  \BibitemOpen
  \bibfield  {author} {\bibinfo {author} {\bibfnamefont {L.~M.}\ \bibnamefont
  {Robledo}}, \bibinfo {author} {\bibfnamefont {T.~R.}\ \bibnamefont
  {Rodr{\'{\i}}guez}},\ and\ \bibinfo {author} {\bibfnamefont {R.~R.}\
  \bibnamefont {Rodr{\'{\i}}guez-Guzm{\'{a}}n}},\ }\href
  {https://doi.org/10.1088/1361-6471/aadebd} {\bibfield  {journal} {\bibinfo
  {journal} {J. Phys. G: Nucl. Part.}\ }\textbf {\bibinfo {volume} {46}},\
  \bibinfo {pages} {013001} (\bibinfo {year} {2018})}\BibitemShut {NoStop}%
\bibitem [{\citenamefont {Shinohara}\ \emph {et~al.}(2006)\citenamefont
  {Shinohara}, \citenamefont {Ohta}, \citenamefont {Nakatsukasa},\ and\
  \citenamefont {Yabana}}]{SO06}%
  \BibitemOpen
  \bibfield  {author} {\bibinfo {author} {\bibfnamefont {S.}~\bibnamefont
  {Shinohara}}, \bibinfo {author} {\bibfnamefont {H.}~\bibnamefont {Ohta}},
  \bibinfo {author} {\bibfnamefont {T.}~\bibnamefont {Nakatsukasa}},\ and\
  \bibinfo {author} {\bibfnamefont {K.}~\bibnamefont {Yabana}},\ }\href
  {https://doi.org/10.1103/PhysRevC.74.054315} {\bibfield  {journal} {\bibinfo
  {journal} {Phys. Rev. C}\ }\textbf {\bibinfo {volume} {74}},\ \bibinfo
  {pages} {054315} (\bibinfo {year} {2006})}\BibitemShut {NoStop}%
\bibitem [{\citenamefont {Bender}\ and\ \citenamefont {Heenen}(2008)}]{BH08}%
  \BibitemOpen
  \bibfield  {author} {\bibinfo {author} {\bibfnamefont {M.}~\bibnamefont
  {Bender}}\ and\ \bibinfo {author} {\bibfnamefont {P.-H.}\ \bibnamefont
  {Heenen}},\ }\href {https://doi.org/10.1103/PhysRevC.78.024309} {\bibfield
  {journal} {\bibinfo  {journal} {Phys. Rev. C}\ }\textbf {\bibinfo {volume}
  {78}},\ \bibinfo {pages} {024309} (\bibinfo {year} {2008})}\BibitemShut
  {NoStop}%
\bibitem [{\citenamefont {Yao}\ \emph {et~al.}(2010)\citenamefont {Yao},
  \citenamefont {Meng}, \citenamefont {Ring},\ and\ \citenamefont
  {Vretenar}}]{YM10}%
  \BibitemOpen
  \bibfield  {author} {\bibinfo {author} {\bibfnamefont {J.~M.}\ \bibnamefont
  {Yao}}, \bibinfo {author} {\bibfnamefont {J.}~\bibnamefont {Meng}}, \bibinfo
  {author} {\bibfnamefont {P.}~\bibnamefont {Ring}},\ and\ \bibinfo {author}
  {\bibfnamefont {D.}~\bibnamefont {Vretenar}},\ }\href
  {https://doi.org/10.1103/PhysRevC.81.044311} {\bibfield  {journal} {\bibinfo
  {journal} {Phys. Rev. C}\ }\textbf {\bibinfo {volume} {81}},\ \bibinfo
  {pages} {044311} (\bibinfo {year} {2010})}\BibitemShut {NoStop}%
\bibitem [{\citenamefont {Rodr\'{\i}guez}\ and\ \citenamefont
  {Egido}(2011{\natexlab{a}})}]{RT11}%
  \BibitemOpen
  \bibfield  {author} {\bibinfo {author} {\bibfnamefont {T.~R.}\ \bibnamefont
  {Rodr\'{\i}guez}}\ and\ \bibinfo {author} {\bibfnamefont {J.~L.}\
  \bibnamefont {Egido}},\ }\href {https://doi.org/10.1103/PhysRevC.84.051307}
  {\bibfield  {journal} {\bibinfo  {journal} {Phys. Rev. C}\ }\textbf {\bibinfo
  {volume} {84}},\ \bibinfo {pages} {051307} (\bibinfo {year}
  {2011}{\natexlab{a}})}\BibitemShut {NoStop}%
\bibitem [{\citenamefont {Rodríguez}\ and\ \citenamefont
  {Egido}(2011)}]{TR11}%
  \BibitemOpen
  \bibfield  {author} {\bibinfo {author} {\bibfnamefont {T.~R.}\ \bibnamefont
  {Rodríguez}}\ and\ \bibinfo {author} {\bibfnamefont {J.~L.}\ \bibnamefont
  {Egido}},\ }\href
  {https://doi.org/https://doi.org/10.1016/j.physletb.2011.10.003} {\bibfield
  {journal} {\bibinfo  {journal} {Phys. Lett. B}\ }\textbf {\bibinfo {volume}
  {705}},\ \bibinfo {pages} {255 } (\bibinfo {year} {2011})}\BibitemShut
  {NoStop}%
\bibitem [{\citenamefont {Rodr\'{\i}guez}\ and\ \citenamefont
  {Egido}(2011{\natexlab{b}})}]{RE11}%
  \BibitemOpen
  \bibfield  {author} {\bibinfo {author} {\bibfnamefont {T.~R.}\ \bibnamefont
  {Rodr\'{\i}guez}}\ and\ \bibinfo {author} {\bibfnamefont {J.~L.}\
  \bibnamefont {Egido}},\ }\href {https://doi.org/10.1103/PhysRevC.84.051307}
  {\bibfield  {journal} {\bibinfo  {journal} {Phys. Rev. C}\ }\textbf {\bibinfo
  {volume} {84}},\ \bibinfo {pages} {051307} (\bibinfo {year}
  {2011}{\natexlab{b}})}\BibitemShut {NoStop}%
\bibitem [{\citenamefont {Yao}\ \emph {et~al.}(2013)\citenamefont {Yao},
  \citenamefont {Bender},\ and\ \citenamefont {Heenen}}]{YB13}%
  \BibitemOpen
  \bibfield  {author} {\bibinfo {author} {\bibfnamefont {J.~M.}\ \bibnamefont
  {Yao}}, \bibinfo {author} {\bibfnamefont {M.}~\bibnamefont {Bender}},\ and\
  \bibinfo {author} {\bibfnamefont {P.-H.}\ \bibnamefont {Heenen}},\ }\href
  {https://doi.org/10.1103/PhysRevC.87.034322} {\bibfield  {journal} {\bibinfo
  {journal} {Phys. Rev. C}\ }\textbf {\bibinfo {volume} {87}},\ \bibinfo
  {pages} {034322} (\bibinfo {year} {2013})}\BibitemShut {NoStop}%
\bibitem [{\citenamefont {Fukuoka}\ \emph {et~al.}(2013)\citenamefont
  {Fukuoka}, \citenamefont {Shinohara}, \citenamefont {Funaki}, \citenamefont
  {Nakatsukasa},\ and\ \citenamefont {Yabana}}]{FS13}%
  \BibitemOpen
  \bibfield  {author} {\bibinfo {author} {\bibfnamefont {Y.}~\bibnamefont
  {Fukuoka}}, \bibinfo {author} {\bibfnamefont {S.}~\bibnamefont {Shinohara}},
  \bibinfo {author} {\bibfnamefont {Y.}~\bibnamefont {Funaki}}, \bibinfo
  {author} {\bibfnamefont {T.}~\bibnamefont {Nakatsukasa}},\ and\ \bibinfo
  {author} {\bibfnamefont {K.}~\bibnamefont {Yabana}},\ }\href
  {https://doi.org/10.1103/PhysRevC.88.014321} {\bibfield  {journal} {\bibinfo
  {journal} {Phys. Rev. C}\ }\textbf {\bibinfo {volume} {88}},\ \bibinfo
  {pages} {014321} (\bibinfo {year} {2013})}\BibitemShut {NoStop}%
\bibitem [{\citenamefont {Bally}\ \emph {et~al.}(2014)\citenamefont {Bally},
  \citenamefont {Avez}, \citenamefont {Bender},\ and\ \citenamefont
  {Heenen}}]{BA14}%
  \BibitemOpen
  \bibfield  {author} {\bibinfo {author} {\bibfnamefont {B.}~\bibnamefont
  {Bally}}, \bibinfo {author} {\bibfnamefont {B.}~\bibnamefont {Avez}},
  \bibinfo {author} {\bibfnamefont {M.}~\bibnamefont {Bender}},\ and\ \bibinfo
  {author} {\bibfnamefont {P.-H.}\ \bibnamefont {Heenen}},\ }\href
  {https://doi.org/10.1103/PhysRevLett.113.162501} {\bibfield  {journal}
  {\bibinfo  {journal} {Phys. Rev. Lett.}\ }\textbf {\bibinfo {volume} {113}},\
  \bibinfo {pages} {162501} (\bibinfo {year} {2014})}\BibitemShut {NoStop}%
\bibitem [{\citenamefont {Yao}\ \emph {et~al.}(2014)\citenamefont {Yao},
  \citenamefont {Hagino}, \citenamefont {Li}, \citenamefont {Meng},\ and\
  \citenamefont {Ring}}]{Yao2014}%
  \BibitemOpen
  \bibfield  {author} {\bibinfo {author} {\bibfnamefont {J.~M.}\ \bibnamefont
  {Yao}}, \bibinfo {author} {\bibfnamefont {K.}~\bibnamefont {Hagino}},
  \bibinfo {author} {\bibfnamefont {Z.~P.}\ \bibnamefont {Li}}, \bibinfo
  {author} {\bibfnamefont {J.}~\bibnamefont {Meng}},\ and\ \bibinfo {author}
  {\bibfnamefont {P.}~\bibnamefont {Ring}},\ }\href
  {https://doi.org/10.1103/PhysRevC.89.054306} {\bibfield  {journal} {\bibinfo
  {journal} {Phys. Rev. C}\ }\textbf {\bibinfo {volume} {89}},\ \bibinfo
  {pages} {054306} (\bibinfo {year} {2014})}\BibitemShut {NoStop}%
\bibitem [{\citenamefont {Yao}\ \emph {et~al.}(2015)\citenamefont {Yao},
  \citenamefont {Zhou},\ and\ \citenamefont {Li}}]{YZ15}%
  \BibitemOpen
  \bibfield  {author} {\bibinfo {author} {\bibfnamefont {J.~M.}\ \bibnamefont
  {Yao}}, \bibinfo {author} {\bibfnamefont {E.~F.}\ \bibnamefont {Zhou}},\ and\
  \bibinfo {author} {\bibfnamefont {Z.~P.}\ \bibnamefont {Li}},\ }\href
  {https://doi.org/10.1103/PhysRevC.92.041304} {\bibfield  {journal} {\bibinfo
  {journal} {Phys. Rev. C}\ }\textbf {\bibinfo {volume} {92}},\ \bibinfo
  {pages} {041304} (\bibinfo {year} {2015})}\BibitemShut {NoStop}%
\bibitem [{\citenamefont {Egido}\ and\ \citenamefont {Jungclaus}(2020)}]{EJ20}%
  \BibitemOpen
  \bibfield  {author} {\bibinfo {author} {\bibfnamefont {J.~L.}\ \bibnamefont
  {Egido}}\ and\ \bibinfo {author} {\bibfnamefont {A.}~\bibnamefont
  {Jungclaus}},\ }\href {https://doi.org/10.1103/PhysRevLett.125.192504}
  {\bibfield  {journal} {\bibinfo  {journal} {Phys. Rev. Lett.}\ }\textbf
  {\bibinfo {volume} {125}},\ \bibinfo {pages} {192504} (\bibinfo {year}
  {2020})}\BibitemShut {NoStop}%
\bibitem [{\citenamefont {Marumori}\ \emph {et~al.}(1980)\citenamefont
  {Marumori}, \citenamefont {Maskawa}, \citenamefont {Sakata},\ and\
  \citenamefont {Kuriyama}}]{MM80}%
  \BibitemOpen
  \bibfield  {author} {\bibinfo {author} {\bibfnamefont {T.}~\bibnamefont
  {Marumori}}, \bibinfo {author} {\bibfnamefont {T.}~\bibnamefont {Maskawa}},
  \bibinfo {author} {\bibfnamefont {F.}~\bibnamefont {Sakata}},\ and\ \bibinfo
  {author} {\bibfnamefont {A.}~\bibnamefont {Kuriyama}},\ }\href
  {https://doi.org/10.1143/PTP.64.1294} {\bibfield  {journal} {\bibinfo
  {journal} {Prog. Theor. Phys.}\ }\textbf {\bibinfo {volume} {64}},\ \bibinfo
  {pages} {1294} (\bibinfo {year} {1980})}\BibitemShut {NoStop}%
\bibitem [{\citenamefont {Matsuo}(1986)}]{M84}%
  \BibitemOpen
  \bibfield  {author} {\bibinfo {author} {\bibfnamefont {M.}~\bibnamefont
  {Matsuo}},\ }\href {https://doi.org/10.1143/PTP.76.372} {\bibfield  {journal}
  {\bibinfo  {journal} {Prog. Theor. Phys.}\ }\textbf {\bibinfo {volume}
  {76}},\ \bibinfo {pages} {372} (\bibinfo {year} {1986})}\BibitemShut
  {NoStop}%
\bibitem [{\citenamefont {Hohenberg}\ and\ \citenamefont {Kohn}(1964)}]{HK64}%
  \BibitemOpen
  \bibfield  {author} {\bibinfo {author} {\bibfnamefont {P.}~\bibnamefont
  {Hohenberg}}\ and\ \bibinfo {author} {\bibfnamefont {W.}~\bibnamefont
  {Kohn}},\ }\href {https://doi.org/10.1103/PhysRev.136.B864} {\bibfield
  {journal} {\bibinfo  {journal} {Phys. Rev.}\ }\textbf {\bibinfo {volume}
  {136}},\ \bibinfo {pages} {B864} (\bibinfo {year} {1964})}\BibitemShut
  {NoStop}%
\bibitem [{\citenamefont {Kohn}\ and\ \citenamefont {Sham}(1965)}]{KS65}%
  \BibitemOpen
  \bibfield  {author} {\bibinfo {author} {\bibfnamefont {W.}~\bibnamefont
  {Kohn}}\ and\ \bibinfo {author} {\bibfnamefont {L.~J.}\ \bibnamefont
  {Sham}},\ }\href {https://doi.org/10.1103/PhysRev.140.A1133} {\bibfield
  {journal} {\bibinfo  {journal} {Phys. Rev.}\ }\textbf {\bibinfo {volume}
  {140}},\ \bibinfo {pages} {A1133} (\bibinfo {year} {1965})}\BibitemShut
  {NoStop}%
\bibitem [{\citenamefont {Runge}\ and\ \citenamefont {Gross}(1984)}]{RG84}%
  \BibitemOpen
  \bibfield  {author} {\bibinfo {author} {\bibfnamefont {E.}~\bibnamefont
  {Runge}}\ and\ \bibinfo {author} {\bibfnamefont {E.~K.~U.}\ \bibnamefont
  {Gross}},\ }\href {https://doi.org/10.1103/PhysRevLett.52.997} {\bibfield
  {journal} {\bibinfo  {journal} {Phys. Rev. Lett.}\ }\textbf {\bibinfo
  {volume} {52}},\ \bibinfo {pages} {997} (\bibinfo {year} {1984})}\BibitemShut
  {NoStop}%
\bibitem [{\citenamefont {Nakatsukasa}\ \emph {et~al.}(2016)\citenamefont
  {Nakatsukasa}, \citenamefont {Matsuyanagi}, \citenamefont {Matsuo},\ and\
  \citenamefont {Yabana}}]{NM06}%
  \BibitemOpen
  \bibfield  {author} {\bibinfo {author} {\bibfnamefont {T.}~\bibnamefont
  {Nakatsukasa}}, \bibinfo {author} {\bibfnamefont {K.}~\bibnamefont
  {Matsuyanagi}}, \bibinfo {author} {\bibfnamefont {M.}~\bibnamefont
  {Matsuo}},\ and\ \bibinfo {author} {\bibfnamefont {K.}~\bibnamefont
  {Yabana}},\ }\href {https://doi.org/10.1103/RevModPhys.88.045004} {\bibfield
  {journal} {\bibinfo  {journal} {Rev. Mod. Phys.}\ }\textbf {\bibinfo {volume}
  {88}},\ \bibinfo {pages} {045004} (\bibinfo {year} {2016})}\BibitemShut
  {NoStop}%
\bibitem [{\citenamefont {Casida}\ and\ \citenamefont
  {Huix-Rotllant}(2012)}]{CH12}%
  \BibitemOpen
  \bibfield  {author} {\bibinfo {author} {\bibfnamefont {M.}~\bibnamefont
  {Casida}}\ and\ \bibinfo {author} {\bibfnamefont {M.}~\bibnamefont
  {Huix-Rotllant}},\ }\href
  {https://doi.org/10.1146/annurev-physchem-032511-143803} {\bibfield
  {journal} {\bibinfo  {journal} {Ann. Rev. Phys. Chem.}\ }\textbf {\bibinfo
  {volume} {63}},\ \bibinfo {pages} {287} (\bibinfo {year} {2012})},\ \bibinfo
  {note} {pMID: 22242728}\BibitemShut {NoStop}%
\bibitem [{\citenamefont {Peierls}\ and\ \citenamefont
  {Thouless}(1962)}]{PT62}%
  \BibitemOpen
  \bibfield  {author} {\bibinfo {author} {\bibfnamefont {P.}~\bibnamefont
  {Peierls}}\ and\ \bibinfo {author} {\bibfnamefont {D.}~\bibnamefont
  {Thouless}},\ }\href
  {https://doi.org/https://doi.org/10.1016/0029-5582(62)91025-8} {\bibfield
  {journal} {\bibinfo  {journal} {Nucl. Phys.}\ }\textbf {\bibinfo {volume}
  {38}},\ \bibinfo {pages} {154 } (\bibinfo {year} {1962})}\BibitemShut
  {NoStop}%
\bibitem [{\citenamefont {Borrajo}\ \emph {et~al.}(2015)\citenamefont
  {Borrajo}, \citenamefont {Rodríguez},\ and\ \citenamefont {{Luis
  Egido}}}]{BR15}%
  \BibitemOpen
  \bibfield  {author} {\bibinfo {author} {\bibfnamefont {M.}~\bibnamefont
  {Borrajo}}, \bibinfo {author} {\bibfnamefont {T.~R.}\ \bibnamefont
  {Rodríguez}},\ and\ \bibinfo {author} {\bibfnamefont {J.}~\bibnamefont
  {{Luis Egido}}},\ }\href
  {https://doi.org/https://doi.org/10.1016/j.physletb.2015.05.030} {\bibfield
  {journal} {\bibinfo  {journal} {Phys. Lett. B}\ }\textbf {\bibinfo {volume}
  {746}},\ \bibinfo {pages} {341 } (\bibinfo {year} {2015})}\BibitemShut
  {NoStop}%
\bibitem [{\citenamefont {Egido}\ \emph {et~al.}(2016)\citenamefont {Egido},
  \citenamefont {Borrajo},\ and\ \citenamefont {Rodr\'{\i}guez}}]{EB16}%
  \BibitemOpen
  \bibfield  {author} {\bibinfo {author} {\bibfnamefont {J.~L.}\ \bibnamefont
  {Egido}}, \bibinfo {author} {\bibfnamefont {M.}~\bibnamefont {Borrajo}},\
  and\ \bibinfo {author} {\bibfnamefont {T.~R.}\ \bibnamefont
  {Rodr\'{\i}guez}},\ }\href {https://doi.org/10.1103/PhysRevLett.116.052502}
  {\bibfield  {journal} {\bibinfo  {journal} {Phys. Rev. Lett.}\ }\textbf
  {\bibinfo {volume} {116}},\ \bibinfo {pages} {052502} (\bibinfo {year}
  {2016})}\BibitemShut {NoStop}%
\bibitem [{\citenamefont {Rodr\'{\i}guez}(2016)}]{R16}%
  \BibitemOpen
  \bibfield  {author} {\bibinfo {author} {\bibfnamefont {T.~R.}\ \bibnamefont
  {Rodr\'{\i}guez}},\ }\href {https://doi.org/10.1140/epja/i2016-16190-2}
  {\bibfield  {journal} {\bibinfo  {journal} {Eur. Phys. J. A}\ }\textbf
  {\bibinfo {volume} {52}},\ \bibinfo {pages} {190} (\bibinfo {year}
  {2016})}\BibitemShut {NoStop}%
\bibitem [{\citenamefont {Shimada}\ \emph {et~al.}(2016)\citenamefont
  {Shimada}, \citenamefont {Tagami},\ and\ \citenamefont
  {Shimizu}}]{Shimada2016}%
  \BibitemOpen
  \bibfield  {author} {\bibinfo {author} {\bibfnamefont {M.}~\bibnamefont
  {Shimada}}, \bibinfo {author} {\bibfnamefont {S.}~\bibnamefont {Tagami}},\
  and\ \bibinfo {author} {\bibfnamefont {Y.~R.}\ \bibnamefont {Shimizu}},\
  }\href {https://doi.org/10.1103/PhysRevC.93.044317} {\bibfield  {journal}
  {\bibinfo  {journal} {Phys. Rev. C}\ }\textbf {\bibinfo {volume} {93}},\
  \bibinfo {pages} {044317} (\bibinfo {year} {2016})}\BibitemShut {NoStop}%
\bibitem [{\citenamefont {Shimada}\ \emph {et~al.}(2015)\citenamefont
  {Shimada}, \citenamefont {Tagami},\ and\ \citenamefont {Shimizu}}]{ST15}%
  \BibitemOpen
  \bibfield  {author} {\bibinfo {author} {\bibfnamefont {M.}~\bibnamefont
  {Shimada}}, \bibinfo {author} {\bibfnamefont {S.}~\bibnamefont {Tagami}},\
  and\ \bibinfo {author} {\bibfnamefont {Y.~R.}\ \bibnamefont {Shimizu}},\
  }\href {https://doi.org/10.1093/ptep/ptv073} {\bibfield  {journal} {\bibinfo
  {journal} {Prog. Theor. Exp. Phys}\ }\textbf {\bibinfo {volume} {2015}},\
  \bibinfo {pages} {063D02} (\bibinfo {year} {2015})}\BibitemShut {NoStop}%
\bibitem [{\citenamefont {Ushitani}\ \emph {et~al.}(2019)\citenamefont
  {Ushitani}, \citenamefont {Tagami},\ and\ \citenamefont
  {Shimizu}}]{Ushitani2019}%
  \BibitemOpen
  \bibfield  {author} {\bibinfo {author} {\bibfnamefont {M.}~\bibnamefont
  {Ushitani}}, \bibinfo {author} {\bibfnamefont {S.}~\bibnamefont {Tagami}},\
  and\ \bibinfo {author} {\bibfnamefont {Y.~R.}\ \bibnamefont {Shimizu}},\
  }\href {https://doi.org/10.1103/PhysRevC.99.064328} {\bibfield  {journal}
  {\bibinfo  {journal} {Phys. Rev. C}\ }\textbf {\bibinfo {volume} {99}},\
  \bibinfo {pages} {064328} (\bibinfo {year} {2019})}\BibitemShut {NoStop}%
\bibitem [{\citenamefont {Goeke}\ and\ \citenamefont {Reinhard}(1980)}]{GR80}%
  \BibitemOpen
  \bibfield  {author} {\bibinfo {author} {\bibfnamefont {K.}~\bibnamefont
  {Goeke}}\ and\ \bibinfo {author} {\bibfnamefont {P.-G.}\ \bibnamefont
  {Reinhard}},\ }\href
  {https://doi.org/https://doi.org/10.1016/0003-4916(80)90210-9} {\bibfield
  {journal} {\bibinfo  {journal} {Ann. Phys. (N. Y.)}\ }\textbf {\bibinfo
  {volume} {124}},\ \bibinfo {pages} {249 } (\bibinfo {year}
  {1980})}\BibitemShut {NoStop}%
\bibitem [{\citenamefont {Klüpfel}\ \emph {et~al.}(2008)\citenamefont
  {Klüpfel}, \citenamefont {Erler}, \citenamefont {Reinhard},\ and\
  \citenamefont {Maruhn}}]{KE08}%
  \BibitemOpen
  \bibfield  {author} {\bibinfo {author} {\bibfnamefont {P.}~\bibnamefont
  {Klüpfel}}, \bibinfo {author} {\bibfnamefont {J.}~\bibnamefont {Erler}},
  \bibinfo {author} {\bibfnamefont {P.~G.}\ \bibnamefont {Reinhard}},\ and\
  \bibinfo {author} {\bibfnamefont {J.~A.}\ \bibnamefont {Maruhn}},\ }\href
  {https://doi.org/10.1140/epja/i2008-10633-3} {\bibfield  {journal} {\bibinfo
  {journal} {Eur. Phys. J. A}\ }\textbf {\bibinfo {volume} {37}},\ \bibinfo
  {pages} {343} (\bibinfo {year} {2008})}\BibitemShut {NoStop}%
\bibitem [{\citenamefont {Hizawa}\ \emph {et~al.}(2021)\citenamefont {Hizawa},
  \citenamefont {Hagino},\ and\ \citenamefont {Yoshida}}]{HH21}%
  \BibitemOpen
  \bibfield  {author} {\bibinfo {author} {\bibfnamefont {N.}~\bibnamefont
  {Hizawa}}, \bibinfo {author} {\bibfnamefont {K.}~\bibnamefont {Hagino}},\
  and\ \bibinfo {author} {\bibfnamefont {K.}~\bibnamefont {Yoshida}},\ }\href
  {https://doi.org/10.1103/PhysRevC.103.034313} {\bibfield  {journal} {\bibinfo
   {journal} {Phys. Rev. C}\ }\textbf {\bibinfo {volume} {103}},\ \bibinfo
  {pages} {034313} (\bibinfo {year} {2021})}\BibitemShut {NoStop}%
\bibitem [{\citenamefont {Kanada-En'yo}\ \emph {et~al.}(2012)\citenamefont
  {Kanada-En'yo}, \citenamefont {Kimura},\ and\ \citenamefont {Ono}}]{KK12}%
  \BibitemOpen
  \bibfield  {author} {\bibinfo {author} {\bibfnamefont {Y.}~\bibnamefont
  {Kanada-En'yo}}, \bibinfo {author} {\bibfnamefont {M.}~\bibnamefont
  {Kimura}},\ and\ \bibinfo {author} {\bibfnamefont {A.}~\bibnamefont {Ono}},\
  }\href {https://doi.org/10.1093/ptep/pts001} {\bibfield  {journal} {\bibinfo
  {journal} {Prog. Theor. Exp. Phys}\ }\textbf {\bibinfo {volume} {2012}}
  (\bibinfo {year} {2012})}\BibitemShut {NoStop}%
\bibitem [{\citenamefont {Hizawa}()}]{hizawa22}%
  \BibitemOpen
  \bibfield  {author} {\bibinfo {author} {\bibfnamefont {N.}~\bibnamefont
  {Hizawa}},\ }\href@noop {} {}\bibinfo {howpublished} {to be
  published}\BibitemShut {NoStop}%
\bibitem [{\citenamefont {Decharg\'e}\ and\ \citenamefont
  {Gogny}(1980)}]{DG80}%
  \BibitemOpen
  \bibfield  {author} {\bibinfo {author} {\bibfnamefont {J.}~\bibnamefont
  {Decharg\'e}}\ and\ \bibinfo {author} {\bibfnamefont {D.}~\bibnamefont
  {Gogny}},\ }\href {https://doi.org/10.1103/PhysRevC.21.1568} {\bibfield
  {journal} {\bibinfo  {journal} {Phys. Rev. C}\ }\textbf {\bibinfo {volume}
  {21}},\ \bibinfo {pages} {1568} (\bibinfo {year} {1980})}\BibitemShut
  {NoStop}%
\bibitem [{\citenamefont {Berger}\ \emph {et~al.}(1991)\citenamefont {Berger},
  \citenamefont {Girod},\ and\ \citenamefont {Gogny}}]{BG91}%
  \BibitemOpen
  \bibfield  {author} {\bibinfo {author} {\bibfnamefont {J.}~\bibnamefont
  {Berger}}, \bibinfo {author} {\bibfnamefont {M.}~\bibnamefont {Girod}},\ and\
  \bibinfo {author} {\bibfnamefont {D.}~\bibnamefont {Gogny}},\ }\href
  {https://doi.org/https://doi.org/10.1016/0010-4655(91)90263-K} {\bibfield
  {journal} {\bibinfo  {journal} {Comput. Phys. Commun.}\ }\textbf {\bibinfo
  {volume} {63}},\ \bibinfo {pages} {365} (\bibinfo {year} {1991})}\BibitemShut
  {NoStop}%
\bibitem [{\citenamefont {Balian}\ and\ \citenamefont {Brezin}(1969)}]{BB69}%
  \BibitemOpen
  \bibfield  {author} {\bibinfo {author} {\bibfnamefont {R.}~\bibnamefont
  {Balian}}\ and\ \bibinfo {author} {\bibfnamefont {E.}~\bibnamefont
  {Brezin}},\ }\href {https://doi.org/https://doi.org/10.1007/BF02710281}
  {\bibfield  {journal} {\bibinfo  {journal} {Il Nuovo Cimento B (1965-1970)}\
  }\textbf {\bibinfo {volume} {64}},\ \bibinfo {pages} {37} (\bibinfo {year}
  {1969})}\BibitemShut {NoStop}%
\bibitem [{\citenamefont {Bonche}\ \emph {et~al.}(1990)\citenamefont {Bonche},
  \citenamefont {Dobaczewski}, \citenamefont {Flocard}, \citenamefont
  {Heenen},\ and\ \citenamefont {Meyer}}]{BD90}%
  \BibitemOpen
  \bibfield  {author} {\bibinfo {author} {\bibfnamefont {P.}~\bibnamefont
  {Bonche}}, \bibinfo {author} {\bibfnamefont {J.}~\bibnamefont {Dobaczewski}},
  \bibinfo {author} {\bibfnamefont {H.}~\bibnamefont {Flocard}}, \bibinfo
  {author} {\bibfnamefont {P.-H.}\ \bibnamefont {Heenen}},\ and\ \bibinfo
  {author} {\bibfnamefont {J.}~\bibnamefont {Meyer}},\ }\href
  {https://doi.org/https://doi.org/10.1016/0375-9474(90)90062-Q} {\bibfield
  {journal} {\bibinfo  {journal} {Nucl. Phys. A}\ }\textbf {\bibinfo {volume}
  {510}},\ \bibinfo {pages} {466 } (\bibinfo {year} {1990})}\BibitemShut
  {NoStop}%
\bibitem [{\citenamefont {Perez}\ \emph {et~al.}(2017)\citenamefont {Perez},
  \citenamefont {Schunck}, \citenamefont {Lasseri}, \citenamefont {Zhang},\
  and\ \citenamefont {Sarich}}]{HFBTHOver3.0}%
  \BibitemOpen
  \bibfield  {author} {\bibinfo {author} {\bibfnamefont {R.~N.}\ \bibnamefont
  {Perez}}, \bibinfo {author} {\bibfnamefont {N.}~\bibnamefont {Schunck}},
  \bibinfo {author} {\bibfnamefont {R.-D.}\ \bibnamefont {Lasseri}}, \bibinfo
  {author} {\bibfnamefont {C.}~\bibnamefont {Zhang}},\ and\ \bibinfo {author}
  {\bibfnamefont {J.}~\bibnamefont {Sarich}},\ }\href
  {https://doi.org/https://doi.org/10.1016/j.cpc.2017.06.022} {\bibfield
  {journal} {\bibinfo  {journal} {Comput. Phys. Commun.}\ }\textbf {\bibinfo
  {volume} {220}},\ \bibinfo {pages} {363} (\bibinfo {year}
  {2017})}\BibitemShut {NoStop}%
\bibitem [{\citenamefont {Flocard}\ and\ \citenamefont
  {Vautherin}(1976)}]{FV76}%
  \BibitemOpen
  \bibfield  {author} {\bibinfo {author} {\bibfnamefont {H.}~\bibnamefont
  {Flocard}}\ and\ \bibinfo {author} {\bibfnamefont {D.}~\bibnamefont
  {Vautherin}},\ }\href {https://doi.org/10.1016/0375-9474(76)90428-0}
  {\bibfield  {journal} {\bibinfo  {journal} {Nucl. Phys. A}\ }\textbf
  {\bibinfo {volume} {264}},\ \bibinfo {pages} {197} (\bibinfo {year}
  {1976})}\BibitemShut {NoStop}%
\bibitem [{\citenamefont {Johnson}\ \emph {et~al.}(2020)\citenamefont
  {Johnson}, \citenamefont {Luu},\ and\ \citenamefont {Lu}}]{JL20}%
  \BibitemOpen
  \bibfield  {author} {\bibinfo {author} {\bibfnamefont {C.~W.}\ \bibnamefont
  {Johnson}}, \bibinfo {author} {\bibfnamefont {K.~A.}\ \bibnamefont {Luu}},\
  and\ \bibinfo {author} {\bibfnamefont {Y.}~\bibnamefont {Lu}},\ }\href
  {https://doi.org/10.1088/1361-6471/abacda} {\bibfield  {journal} {\bibinfo
  {journal} {J. Phys. G: Nucl. Part.}\ }\textbf {\bibinfo {volume} {47}},\
  \bibinfo {pages} {105107} (\bibinfo {year} {2020})}\BibitemShut {NoStop}%
\end{thebibliography}%

\end{document}